\documentclass[letterpaper]{article} 
\usepackage{aaai2026}  
\usepackage{times}  
\usepackage{helvet}  
\usepackage{courier}  
\usepackage[hyphens]{url}  
\usepackage{graphicx} 
\usepackage{natbib}  
\usepackage{caption} 
\usepackage{algorithm}
\usepackage{algorithmic}

\usepackage[]{mdframed}
\usepackage{tcolorbox}
\usepackage{newfloat}
\usepackage{listings}
\DeclareCaptionStyle{ruled}{labelfont=normalfont,labelsep=colon,strut=off} 
\floatstyle{ruled}
\newfloat{listing}{tb}{lst}{}
\floatname{listing}{Listing}
\nocopyright 

\newcommand\blfootnote[1]{%
  \begingroup
  \renewcommand\thefootnote{}\footnote{#1}%
  \addtocounter{footnote}{-1}%
  \endgroup
}

\title{Visibility Allocation Systems: 
How Algorithmic Design Shapes Online Visibility and Societal Outcomes
\blfootnote{This work was supported as a part of NCCR Automation, a National Centre of Competence in Research, funded by the Swiss National Science Foundation (grants number 51NF40\_225155 and 215354). It was also supported by the Kone Foundation.}}

\author{
    Stefania Ionescu$^1$,
    Robin Forsberg$^2$, 
    Elsa Lichtenegger$^2$, 
    Salima Jaoua$^2$, 
    Kshitijaa Jaglan$^2$,\\
    Florian Dörfler$^1$,
    Anikó Hannák$^2$
}
\affiliations{

    $^1$ ETH Zurich, 
    $^2$ University of Zürich\\
}

\begin{document}

\maketitle

\begin{abstract}
Throughout application domains, we now rely extensively on algorithmic systems to engage with ever-expanding datasets of information.
Despite their benefits, these systems are often complex (comprising of many intricate tools, e.g., moderation, recommender systems, prediction models), of unknown structure (due to the lack of accompanying documentation), and having hard-to-predict yet potentially severe downstream consequences (due to the extensive use, systematic enactment of existing errors, and many comprising feedback loops). As such, understanding and evaluating these systems as a whole remains a challenge for both researchers and legislators.
To aid ongoing efforts, we introduce a formal framework for such \emph{visibility allocation systems} (VASs) which we define as (semi-)automated systems deciding which (processed) data to present a human user with. We review typical tools comprising VASs and define the associated computational problems they solve. By doing so, VASs can be decomposed into sub-processes and illustrated via data flow diagrams. Moreover, we survey metrics for evaluating VASs throughout the pipeline, thus aiding system diagnostics. Using forecasting-based recommendations in school choice as a case study, we demonstrate how our framework can support VAS evaluation.
We also discuss how our framework can support ongoing AI-legislative efforts to locate obligations, quantify systemic risks, and enable adaptive compliance.
\end{abstract}



\section{Introduction}
Throughout application domains, we have ever-increasing amounts of data at our disposal. To avoid data smog (also referred to as information overload), an algorithmic layer makes a selection for the user~\cite{shenk1998data, bawden2020information}. Search engines return lists of articles related to a given query, social media platforms recommend engaging content, online marketplaces provide filters and sorting tools to support satisfactory purchases, while automatic CV assessment tools make a first screening of job candidates. These are just a few examples of technical systems embedded in our society for deciding which \textit{alternatives} (documents, videos, posts, opinions, products, job candidates) to make visible to a given human user. In this work, we will call them \textit{visibility allocation systems} (VASs).

VASs have important societal effects. While some of these effects are positive (e.g., increased access to information, low cost of sharing decision-support systems), others are negative. Extensive reliance on personalized recommendations can lead to convergence in taste~\cite{pajkovic2021algorithms, beer2013algorithms}, echo chambers, and filter bubbles~\cite{nguyen2014exploring, geschke2019triple}. Machine learning models used in high-stakes decision-making processes can suffer from and even amplify biases because of unsuitable training sets~\cite{barocas2016big, feldman2015certifying} or measurements~\cite{jacobs2021measurement, friedler2019comparative}. More generally, systems of machines influence culture evolution as they generate, select, transmit, and alter cultural artifacts~\cite{brinkmann2023machine}. Due to all their influence on society, both researchers and legislators underline the importance of understanding the long-term effects of such processes~\cite{narayanan2023understanding, brinkmann2023machine}.

However, evaluating the long-term effects of VASs is challenging for several reasons. First, these systems are complex, consisting of multiple intricate tools (e.g., moderation, recommendation systems, prediction models), which are often developed and evaluated in isolation. Second, the deployment of these technical systems within the real world leads to a feedback loop between VASs and people, which creates an additional layer of complexity.
Third, the lack of transparency of VAS architecture leaves many unknowns, such as the current implementation of comprising tools and the interaction between them. Moreover, VASs change frequently, leaving ambiguous which version is being evaluated and whether takeaways still apply to current versions. All these are just a few examples of related challenges.

Perhaps the biggest challenge in understanding and evaluating VASs is the large number of relevant research fields and methodologies. Computer science focuses on the development of efficient algorithms, Economics on the analysis of strategic user behavior, Psychology on understanding individual human behavior, Sociology on studying the emergence of groups and collective behavior, Law, Ethics, and Philosophy on identifying and deploying desired properties of systems. The emergence of fields such as Human-Computer Interaction, Algorithmic Fairness, and Economics and Computation helps bridge between communities, but cannot, of course, diminish the diversity of relevant expertise. As a result, various methods and procedures offer complementary insight into the study of VASs, e.g., complexity theory, formal verification, algorithmic auditing, experiments, surveys, data analysis, and agent-based modeling.

The goal of our work is to unify the language surrounding (semi)automated systems that decide which (processed) data to present a human user with. We do so by providing a formal abstraction giving foundational support for both the analysis and design of VASs. Through our formal framework for describing and understanding such systems, we aim to lower the friction of information transfer between domains. Ultimately, we show this makes the study of long-term societal effects easier and supports ongoing legislative efforts.

Doing so leads to several contributions. Namely, we:
\begin{enumerate}
    \item Define the visibility allocation problem and systems.
    \item Formalize the most frequently used tools within VASs (e.g., search, filter, recommendation, moderation). We provide compatible definitions for them, effectively decomposing VASs into sub-processes for a more granular evaluation.
    \item Demonstrate how VASs can be represented by data-flow diagrams (DFDs) and their comprising tools as processes within DFDs. Logs can document systemic changes and learnings.
    Such high-level system representations promote better understanding and increase transparency.
    \item Review both tool-specific and long-term societal metrics relevant to studying VASs. Executable versions of DFDs help connect design choices to their long-term effects.
    \item Using forecasting-based recommendations in the context of school choice as a case study, demonstrate how to operationalize our framework and how our abstraction aids in the evaluation of long-term effects of VASs.
\end{enumerate}

\section{Brief Overview and Context of Related Work\footnote{We include a more extensive literature review in the appendix.}
}

\paragraph{Other (semi)-automated systems.} 
While already existing system definitions bear similarities to the \textit{visibility allocation systems} (VASs), they are overall distinct. \textit{Information access systems} also aim to present users with items that satisfy their information need~\cite{ekstrand2022fairness}, and encompass both \textit{information retrieval} (IR) and \textit{information filtering} (IF)~\cite{belkin1992information, hanani2001information}. Still, they do not include, e.g, machine learning models trained to make predictions based on offline interaction data. While \textit{decision support systems} include such models~\cite{bourgeois2019information}, they do not cover other processes that influence both individuals and society, such as ad auctions~\cite{ali2019discrimination}. This is why prior work resorts to local, ad-hoc terms like `the algorithm'~\cite{grandinetti2023examining}, semi-autonomous algorithmic technologies~\cite{pajkovic2021algorithms}, systems for content processing and propagation~\cite{narayanan2023understanding}. To cover this gap, we introduce VAS for referring to all technical systems that decide which alternatives are visible to users. Similar to this prior work, we define VASs based on the computational problem they address~\cite{ekstrand2022fairness}; additionally, we provide definitions for the problems associated with the comprising tools.

\paragraph{Societal impact of VASs.} Most studies focus on the societal impact of either specific tools or specific instances of platform design. For example, machine learning models can learn biased representations from training data~\citep{Jiawei2023}, leading to algorithmic biases and the reinforcement of existing societal inequalities~\citep{oneil2016}, effects often amplified through feedback loops \citep{pagan2023, mansoury2020}. At a platform level, studies on social media show they can contribute to the spread of misinformation~\citep{Vicario2016,Vosoughi2018}, create echo chambers~\citep{Cinelli2021, delvicario2016echo}, and lead to polarization~\citep{Garimella_Weber_2017}. Platform changes can reshape user behavior: e.g., the introduction of the feature 'People You May Know' on Facebook led to a drastic increase in triangles and new connections~\cite{Zignani_2014}.
Our VAS framework supports documenting such systemic changes and learnings, as well as connecting specific design choices to their societal impact and transferring learnings across application domains.

\paragraph{Relevant research fields and methodologies.} More work aiming to integrate multidisciplinary perspectives is needed to ensure stakeholder-aligned system design~\cite{Langer2021, Lenders2024}. Computer science and engineering provide the technical foundation through algorithm development and audits, focusing on system architecture and optimization ~\cite{Ricci2011, Metaxa2021, soares2024review}. Yet, it does not give a complete picture. Fairness, transparency, and accountability in AI, for example, need to be grounded in the broader legal, social, and ethical contexts~\cite{Starke2025, Cheong2024}. Mitigating structural inequalities in algorithmic decision-making requires additional insight from social sciences~\cite{Gerdon2022}. Moreover, policy and system design changes can alter individual behavior. To this end, economics uses game theory, mechanism design, and agent-based models to investigate changes in incentives that could lead to unwanted strategic behavior~\cite{Varian2007}, while behavioral economics and psychology examine bounded rationality and contextual influences on individual decision-making~\cite{Kahneman2011, Goldstein2008, Arnott2019}.
By providing harmonized terminology with shared representations and definitions, we aim to facilitate cross-disciplinary understanding.
Through structured diagrams and tables, our framework supports a unified tracking of systemic changes, new algorithmic developments, and findings from data analysis and experiments carried out on particular VASs or demographics; this helps to quickly integrate and transfer knowledge across domains.


\paragraph{Legislation.} As both the tools comprising VASs and specific implementations of VASs are known to have important societal consequences, they are also the subject of current regulatory efforts.  On AI Systems alone, there are several key pieces of legislation: EU's Artificial Intelligence Act \cite{EU_AI_Act_2024}, Digital Services Act \cite{EU_Digital_Services_Act_2022}, and the General Data Protection Regulation \cite{GDPR}. Additionally, UNESCO's Recommendation on the Ethics of AI \cite{UNESCO_AI_Ethics_2021} as well as US Executive Order 14110 \cite{US_Executive_Order_14110_2023} (revoked in January 2025) have served as influential normative reference points. These aim to provide harmonised rules regarding the development and deployment of AI systems, promote transparency (a shared understanding among market operators, public awareness), identify and regulate high-risk systems. Our framework intends to support these efforts by providing a harmonized language, suggesting structured diagrams with accompanying documentation, and using these to more easily integrate learnings (e.g., into executable models) and identify knowledge gaps. We demonstrate how to use this framework to evaluate high-risk systems in one of the critical sectors, education~ \cite{EU_Digital_Services_Act_2022}.

\section{Fundamentals of Visibility Allocation Systems}

\paragraph{VAS.} We call a (semi-)automated system deciding which (processed) data to present to a given human user a visibility allocation system (VAS). VASs, thus, provide a way of deciding which user sees what, i.e. a way of allocating visibility over a set of alternatives. 
More precisely, we define the \textit{visibility allocation problem} as follows:
\begin{quote}
\emph{Given a dataset of alternatives and inputs from stakeholders, present a user with a structured subset (from the space) of alternatives satisfying the needs and requirements of the stakeholders.}
\end{quote}
Above, \textit{structured subset} refers to a subset of alternatives 
arranged for display in a Graphical User Interface; this rendering aspect is important to include as it can influence both accuracy and fairness \cite{chen2011users, beel2021unreasonable, ekstrand2022fairness}. Stakeholders can provide \textit{input} both explicitly (e.g., likes, ratings, filter criteria) and implicitly (e.g., via clicks, watch time)~\cite{schmidt2000implicit}. 
We opt for \textit{alternatives} instead of \textit{items} for two main reasons. First, to emphasize the term can not only refer to items in the classical sense (e.g., products, documents, news, or videos), but also to people (e.g., content creators, job applicants, students, or refugees). Second, to also capture alternatives in a continuous space, such as opinions. Moreover, we say the output can be \textit{from the space of alternatives} to account for more recent tools such as summarization techniques and large language models which can generate new texts that fall outside the dataset of existing documents. 

\paragraph{Tools.}
VASs are composed of a variety of algorithmic tools, 
which share a common operational logic: they take in a set of alternatives, apply some transformation or evaluation, and output a refined subset or reordered version of those alternatives. 
Each tool can be understood in terms of three key dimensions: input, design choices, and output. Input refers to the data a tool processes, typically a set of alternatives, and any user-specified parameters, such as filters, queries, or sorting preferences. Design choices are the internal mechanisms or policies that guide how inputs are handled. These include algorithms, evaluators, or rules, which vary depending on the tool’s function. Output is the result produced after processing, which may be, e.g., a filtered subset, a ranked list, or a classification decision. Although these dimensions are shared across tools, their specific characteristics differ, reflecting the distinct roles each tool plays in the visibility pipeline. In what follows we present two specific tools and the problems they address.

\textbf{Filter} tools select a subset of alternatives based on specified conditions. The input typically includes a set of alternatives and, optionally, user-specified filters or classifiers. The design of a filter tool involves defining evaluators which are functions that assign a binary value to each alternative, determining whether it meets the filtering criteria. The output is a reduced set of alternatives that satisfy the selected evaluators, effectively narrowing down what is visible to the user \citep{Markovitch1993}. Thus, the \textit{filter problem} is: 
\begin{quote}
\emph{Given a dataset of alternatives, a set of evaluators, and a user's set of evaluators, present the maximal subset of alternatives matching the user's evaluators.
}
\end{quote}
Filtering tools play an important role across various contexts: for example they are used to design word filters for moderating comments on social media \citep{Jhaver2022}, to control youth access to online content \citep{Mitchell2005, F_example3}, to allow users to systematically filter out alternatives based on predefined criteria on online dating platforms \citep{F_example4}.

\paragraph{Sort} tools reorder alternatives based on a chosen criterion \citep{knuth1998art}. They take as input a set of alternatives, often along with a user-specified sorting preference such as relevance, date, or popularity. The design involves selecting sorting criteria and a sorting algorithm. 
The output is a permutation of the alternatives, arranged so that their values on the chosen criterion follow a defined order (e.g., descending relevance). Formally, they solve the following \textit{sorting problem}:
\begin{quote}
    \emph{Given a dataset of alternatives, a key function assigning a value to every alternative, return a permutation of the alternatives ordered by their respective values.}
\end{quote}
Sorting tools allow users to change the order based on their preferences or needs. For example, online marketplaces allow sorting by price, ratings, or newest arrivals~\cite{Chu2020}. Social media and news platforms offer sorting by chronological order or trending~\cite{Shmargad2020}. By providing specific sorting choices or setting a default one, platforms exert considerable influence over which information users encounter first, thereby shaping user attention and impacting user decisions~\cite{Robertson2018, Gillespie2018, Eslami2015}.

\textbf{Additional tools.} For brevity, we only present filter and sort in the main text. We choose to prioritize these 
since other tools can be viewed as functional compositions of them, often extended with scoring functions. 
Search, for instance, consists of retrieval (that \textbf{filters}, e.g., documents containing at least one of the query's terms) and ranking (that \textbf{sorts} retrieved documents based on their fit with the query, as estimated by a scoring function)~\cite{nakamura2019anatomy}. Other examples are recommender systems, moderation, and forecasting tools. Their added complexity brings important differences, such as more intricate design choices and societal effects. We define and discuss them in the appendix.

While we consider these tools to be the primarily used ones, the list is far from exhaustive. 
Ad auctions~\cite{Petropoulos_2022} and query suggestions~\cite{Noble2018} are two other examples of tools that affect what is visible to users. 
Moreover, increasingly many people use large language models (LLMs) to inform themselves on a variety of topics~\cite{eysenbach2023role, labadze2023role}.
The recent large-scale adoption of LLMs demonstrates how new visibility allocation tools continue to emerge and evolve. This makes providing a full list of tools both unfeasible and not useful in the long run. Instead, our hope is the tools presented above can serve as examples of how to formulate other visibility allocation tools as data transformers and visibility selectors. As we will demonstrate in the upcoming sections, viewing them as such can facilitate an analysis of the system as a whole, connect design choices with long-term effects, and transfer learnings about these effects between the different tools.
Ultimately, we believe \textit{visibility allocation system} is a needed terminology to comprise the tools commonly used nowadays for viewing information, one which we hope will support the research in the field in a similar way as \textit{information retrieval} and \textit{information filtering}~\cite{belkin1992information, hanani2001information} did for their respective sets of tools.

\section{Illustrating and Evaluating VASs}
The main difficulty in analyzing VASs is connecting tool-level design choices with long-term effects. On the one hand, stakeholders are primarily interested in the long-term effects of VASs: companies in their future revenues, end-users in their over-time satisfaction with the shown alternatives, society and legislators in how they affect societal outcomes. However, developing VASs ultimately requires decisions about the design of each particular tool, which are difficult to connect with the resulting long-term effects of a VAS as a whole. The multitude of stakeholders with often non-aligning goals adds to the complexity of analyzing alternative designs of VASs and choosing between them.

To this end, this section introduces and adapts data-flow diagrams (DFDs) for abstracting and representing VASs within the social context they are deployed. It also reviews commonly used metrics, existing challenges in evaluating VAS-based sociotechnical systems, and how our integrated framework of tool definitions and VAS representation supports connecting design choices to long-term effects.

\subsection{Data Flow Diagrams for Sociotechnical Systems}
\begin{figure*}[t]
\centering
\includegraphics[width=2\columnwidth]{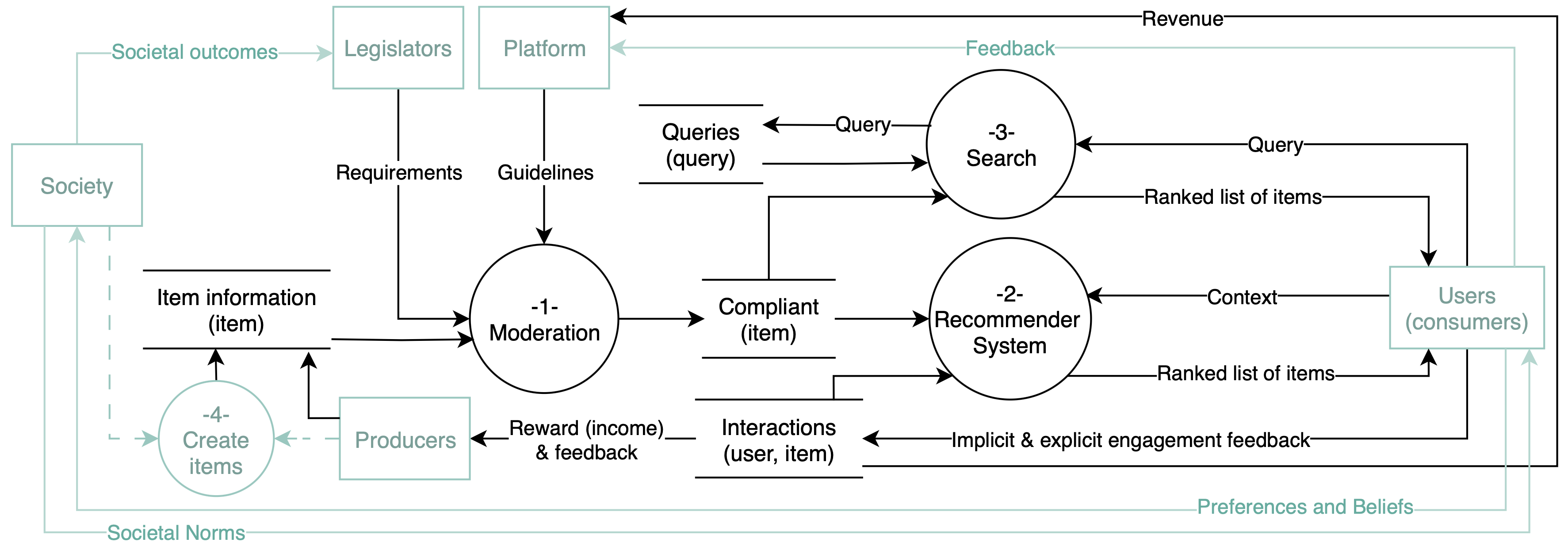} 
\caption{Data-flow diagram for a platform with user-generated content. Dark (black) components are part of the Visibility Allocation System (VAS) with three tools: moderation, recommender system, and search. Light (green) components are external to the VAS. The diagram includes: stakeholders (rectangles), processes (circles), information added to the dataset (open-ended rectangles) indexed by key (annotated in brackets), data flows (full arrows), and participation in a process (dotted arrows).}
\label{fig:mod+rs+search}
\end{figure*}

Data flow diagrams (DFDs) are one of the fundamental structured methodologies for the analysis and design of information systems~\cite{yourdon1979structured,demarco1979structured, siau2022information}. They are graphical representations showing the flow of data throughout a system. Processes (represented by circles) take incoming data and transform it, thus producing new data. Besides other processes, the data can originate or go to datastores (rectangles without vertical sides) or external entities, such as users and producers (rectangles). Arrows capture the direction of data flow and can be annotated with the data being transferred. Figure~\ref{fig:mod+rs+search} shows these elements for a social media platform VAS comprising of moderation, search, and a recommender system.


To adapt DFDs for the study of sociotechnical systems, we make two changes. First, we argue for considering the comprising tools (reviewed in the previous section) as the main processes. This departs from the original focus of DFDs on modeling process structures for business, many of which have limited societal effects~\cite{falkenberg1991understanding}, e.g., \textit{recording payment}, \textit{producing payment confirmation}, and \textit{account creation}~\cite{demarco1979structured, cheema2023natural}. Second, external entities (stakeholders) are no longer simple sources or sinks that are outside the domain of our study~\cite{demarco1979structured}, but could be an active part of processes. For example, when a new YouTube video (item) is created, the content creator (producer) might interview other people, discuss news, or events in a specific cultural context. Thus, the creation of videos becomes a process in which both the producer and society take part. We represent such relationships from \textit{producers} and \textit{society} to the process of \textit{content creation} by dotted arrows. Dotted arrows, thus, represent the relation of being part of a process, while full arrows represent the transfer of data to a process. The appendix further discusses the choice of DFDs and the changes we made.

\subsection{The Impact of VASs on Different Stakeholders}

\begin{figure}[t]
\centering
\includegraphics[width=0.83\columnwidth]{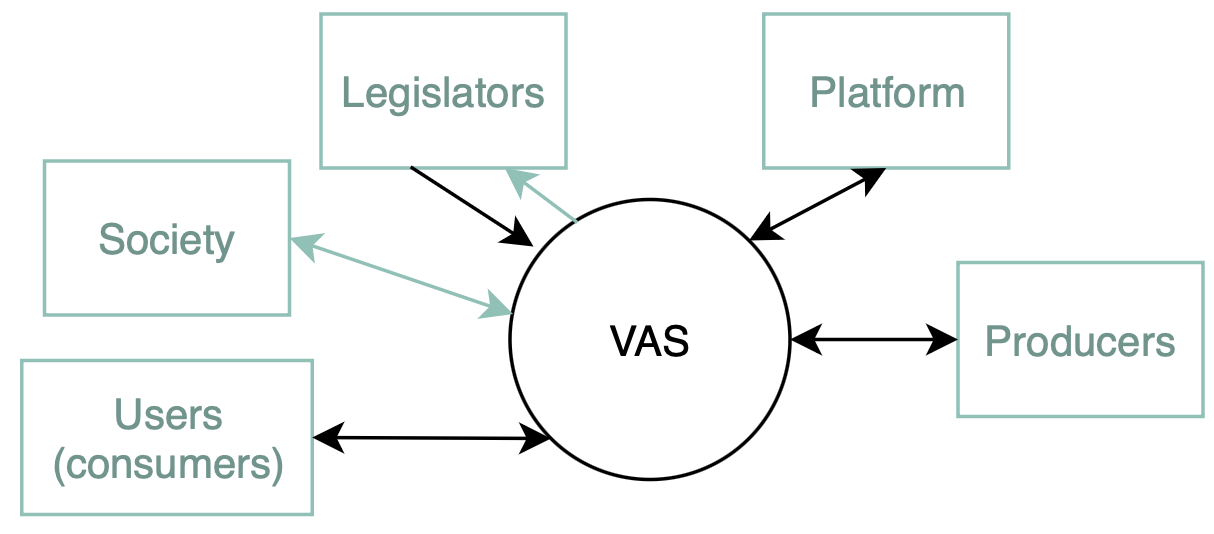} 
\caption{The context diagram showing the VAS for user-generated content in Figure~\ref{fig:mod+rs+search} at the highest level of abstraction (level 0). Dark (black) arrows show direct impact, i.e., via direct information transfer. Light (green) arrows show indirect impact, i.e., via VAS-external influence.}
\label{fig:dfd_l0}
\end{figure}

A key role of dataflow diagrams (DFDs) is to represent the stakeholders involved within a sociotechnical system together with their role and interactions within this system. In fact, context diagrams (DFDs at the highest level of abstraction) just illustrate the VAS and the stakeholders interacting with it~\cite{davis2019data}. Figure~\ref{fig:dfd_l0} provides an example of a context diagram for the VAS introduced earlier. It succinctly shows the main stakeholders, and how they influence and are influenced by the VAS. Some of these influences are direct (e.g., consumers directly provide queries to the VAS) while others are indirect (e.g., society changes based on the updates in preferences and beliefs of individuals). 
Naturally, distinct stakeholders are interested in different aspects affected by VASs. As such, a complete analysis of VASs accounts for metrics of interest to the various stakeholders. 

\emph{Users}, the primary consumers of VASs, have limited control over the mechanisms producing the alternatives they see. They provide, however, input (e.g., via clicks, queries, and ratings) which is fed back into the system, ultimately influencing future VAS outputs~\cite{Rieder2018, Eslami2015}.
Users may prioritize metrics related to perceived service quality and cost-effectiveness, such as recommendation relevance and user satisfaction scores.
\emph{Producers} design, implement, and supply content to VASs and rely on visibility to reach audiences, drive engagement, and generate revenue. While they can shape visibility allocation through strategic behavior, they are largely constrained to platform policies and shifting algorithmic logic~\cite{Gillespie2018, Caplan2020, Bucher2012}.
They focus on engagement metrics like clicks, conversion rates, and overall visibility of their content, which directly impact their revenue and reach~\cite{salganik2006experimental, pagan2021meritocratic, ionescu2023group}. \emph{Platforms} provide the infrastructure and usually emphasize system performance metrics such as accuracy, F1 scores, user satisfaction, usability and profitability~\cite{Bucher2012, Plantin2018}. 

Models could consider \emph{additional stakeholders} affected by the broader societal and ecological impacts of VASs, such as the general public, regulators, and the natural environment. 
We leave details to the appendix, but emphasize DFDs capture paths between processes and various stakeholders and can thus support analysis that goes beyond user- and platform-centric metrics (e.g., considering societal risks, environmental costs, broader patterns of knowledge diffusion).

\subsection{From Design-choices to Long-term Evaluation}



For operational reasons, most of the system evaluation is carried out at a tool level.
For instance, when filtering out undesirable comments on social media platforms, developers and researchers measure content creator satisfaction and accuracy~\cite{Jhaver2022, Sjosten2020}. In sorting, we measure the performance of algorithms through time complexity and stability~\cite{al2013review}. Search engineers, on the other hand, might be concerned with search satisfaction metrics like precision or completeness, document relevance and clicks
~\cite{Chen2017}. 
Recommendation systems use various performance metrics like precision and recall~\cite{rs_2, abdollahpouri2020multistakeholder}.

While these metrics are vital for evaluating the accuracy and reliability of tools within VASs, they fall short of capturing the time dimension and system evolution. 
For example, well-received but biased filter criteria on online dating platforms could lead to racial homogamy, ultimately exacerbating income inequalities~\cite{F_example4}. Surveys and experiments help identify such issues, leading to social theories which can ultimately be tested with agent-based models and simulations~\cite{anderson2014political, ionescu2021agent, zhang2025image}. More generally, many societal relevant metrics, like fairness, are not static and require simulations for connecting low-level design choices with higher-level long-term effects~\cite{d2020fairness}.

To further support interdisciplinary analysis connecting design choices to long-term effects, we suggest accompanying DFDs by logs that document chronological design changes of VASs and findings about specific processes. Doing so can help in several ways. For example, it can help auditing studies shortlist alternative features they could vary and be precise about which VAS design they investigate. It also supports agent-based models overview and integrate the most recent findings, thus producing more realistic and insightful models. Notably, it aligns with ongoing legislative efforts aiming to provide documentation standards for continuous monitoring and increased transparency~\cite{EU_AI_Act_2024}.

\section{Example: VAS for Matching Markets}
Next, we demonstrate how to apply our framework to one specific example, namely school choice. 
For this example, we show how our framework supports a holistic system monitoring and analysis, as well as identifying overlooked phenomena. We also discuss how learnings extend beyond this specific case study.

\subsection{Constructing the Diagram}
In large cities, many new middle-school students arrive every year to be assigned to one of the several existing high schools. This is a difficult task that requires taking into account not only the preferences of students but also several other signals and constraints (e.g., results of students, commute time, siblings already studying at one high school). This is why many governments use a centralized matching mechanism, i.e., students submit their preference (ranking of high schools), and the mechanism returns an allocation (matching each student to one of the high schools)~\cite{abdulkadirouglu2005boston, correa2019school}. 
Because it is difficult for students to decide which high 
school would be best for them, prior efforts suggested leveraging past data, e.g.,
to predict students' development at different schools based on prior students' evolution~\cite{wilson2009smartchoice}.

\begin{figure}[t]
\centering
\includegraphics[width=0.98\columnwidth]{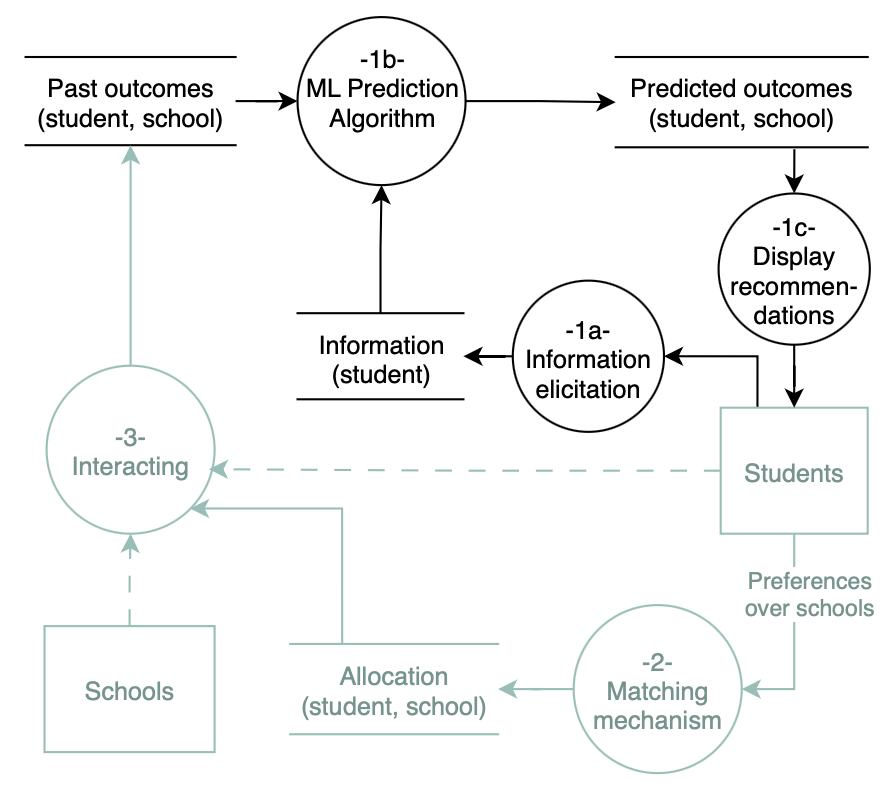} 
\caption{A data-flow diagram for school choice. Dark (black) components are part of the VAS. Light (green) components are external to the VAS. The diagram includes: stakeholders (rectangles), processes (circles), information added to the dataset (open-ended rectangle) indexed by key (annotated in brackets), data flows (full arrows), and participation in a process (dotted arrows).
}
\label{fig:rs+mm}
\end{figure}

We can depict this sociotechnical system using a data-flow diagram (Figure~\ref{fig:rs+mm}). This system has two main stakeholders, namely schools and students. 
The data store keeps track of relevant information about the students, including historical outcomes (e.g., final exam/SAT results) of students who attended courses at different high schools.
The VAS consists of three processes: (1a) an information elicitation aimed to gather students' information, (1b) an ML algorithm that, given a dataset of past outcomes of students who attended courses at different high schools, builds a model that predicts outcomes for all new student-highschool pairs, 
and (1c) a way of presenting the alternative schools to students based on these predictions. There are also two processes external to the VAS: (2)
a matching mechanism that makes an allocation for a given set of preferences and (3) the student-school interactions
which produce new data.

\subsection{Supporting System Evaluation}
\emph{Identifying impacted stakeholders.} A first benefit from using a DFD is identifying the stakeholders (agents) and how they interact with the VAS. In our case, these stakeholders are students and schools. 
As such, the dataset stores important information about these agents at different points in the system, e.g., past and predicted outcomes. Keys (i.e., unique identifiers for each agent) help track results across time and processes.
Ultimately, this supports a multistakeholder evaluation of the sociotechnical system~\cite{abdollahpouri2020multistakeholder}.

\emph{Studying the processes in isolation.} The first layer of analysis is to study processes in isolation (i.e., by analyzing the way they transform data adjacent in the diagram), which is the focus of most prior work.
For instance, we know predictions (process 1b) vary in their levels of accuracy~\cite{mcnee2006being} and bias~\cite{feldman2015certifying}. The design and reputation of the VAS (process 1) can affect the trust of users in the system, and, as a result, the impact of its recommendations on the final preferences of students~\cite{dietvorst2018overcoming}. 
Specifically for the school-choice VAS, surveys show most parents were very satisfied with the provided support~\cite{wilson2009smartchoice}, thus potentially suggesting high levels of trust in the system.
The choice of matching mechanisms (process 2) can lead to different properties of the resulting allocation,
such as strategy-proofness and stability.
Table~\ref{table:processes} summarizes these learnings. Next, we show how these data-based studies support a holistic analysis of the sociotechnical system.

\begin{table}[t]
\centering
\renewcommand{\arraystretch}{1.35}
\small
\begin{tabular}{p{0.5cm}|p{7cm}}
    Pr. & Learning \\
    \hline
    \hline
    1b & One of the main goals is to improve the accuracy of predictions~\cite{mcnee2006being}. \\
    & Algorithms may have biases; e.g., disparate impact (different rates of outcomes depending on user demographics)~\cite{feldman2015certifying, ekstrand2022fairness}.
    \\ \hline
    1 & Algorithmic systems face different levels of users' trust and algorithmic aversion/appreciation (i.e., users' reluctance/willingness to rely on algorithmic as opposed to human advice)~\cite{dietvorst2018overcoming, castelo2019task, jussupow2020we}. \\ 
    & For school choice, surveys show parents were satisfied with the provided support~\cite{wilson2009smartchoice}.\\ \hline
    2 & Prior work studied different matching mechanisms (e.g., serial dictatorship, Boston, Deferred Acceptance). Only some are strategy-proof (i.e., students achieve the best outcome when providing their true preferences)~\cite{abdulkadiroglu2006changing}. \\ \hline
\end{tabular}
\caption{Documents learnings from prior work on the processes (Pr.) depicted in Figure~\ref{fig:rs+mm}.}
\label{table:processes}
\end{table}

\emph{Analyzing sociotechnical systems as a whole.} While literature largely focuses on studying the process in isolation, DFDs can offer a bird's eye perspective. A direct analysis of the DFD can reveal new feedback loops, i.e., cycles in the graph showing how datasets both influence and are influenced by processes. Figure~\ref{fig:rs+mm} shows that past outcomes influence recommendations (predicted outcomes), which affect students' preferences and allocations, which impact the next round of outcomes, thus closing the loop. 
The diagram shows that both students and schools take part. While this feedback loop is glaringly obvious in the DFD, \citet{ionescu2023strategic} discovered it 
only about 15 years after 
\citet{wilson2009smartchoice} proposed the VAS for school choice. 
Results show this feedback loop can incentivize schools to act strategically, and
exacerbate inequalities in the student population. 
Keeping updated versions of the DFDs could support finding such issues faster.

\emph{Evaluating the effects of different design choices.} 
The information in Table~\ref{table:processes} can be brought together into one model in order to reason about the entire system. For example, by abstracting an economic model, prior work proved that schools have an incentive to interact strategically in order to trigger different predictions and allocations\footnote{I.e., high schools which do not want to be matched to under-performing students (e.g., because such students need more costly support for performing well) can interact strategically (by not providing additional support for such students). This signals the VAS not to recommend this school to future under-performing students which leads to fewer matches between the school and such students in the future. For the under-performing students, however, such strategic interactions immediately lead to worse results. }~\cite{ionescu2023strategic}. An alternative is to create an agent-based model with all the information and implement each process using a programming language, thus building an executable version (i.e., simulator) of the DFD in Figure~\ref{fig:rs+mm}. Studying this digital twin of the original system provides insight into the long-term outcomes (e.g., student inequalities, 
magnitude of utility gains by interacting strategically) under different design choices and market characteristics (e.g., matching mechanisms, accuracy of the prediction model, market competition). 
This demonstrates that using modeling to make DFDs executable can test the long-term effects of sociotechnical systems under alternative system designs.

\emph{Guiding future research questions.} By creating and analyzing DFDs we can identify gaps for future research to fill. In school choice, the diagram showed the presence of the interaction-allocation feedback loop. A review of prior work on the comprising processes (Table~\ref{table:processes}) reveals little is known about strategic interactions (process 3).
While these are often optimal in theory~\cite{ionescu2023strategic}, future work can use surveys (e.g., as done for strategic reporting in course allocations~\cite{budish2012multi}) and experiments (e.g., via games~\cite{chakraborty2023influence}) to assess the extent to which agents would find and use optimal strategic interactions. If the results confirm that strategic interacting is also a problem in practice, future work can investigate how to best redesign the system.
As underlined by \citet{gilbert2005simulation}, modeling and simulations can thus highlight important directions for future research.

\subsection{Documenting System Evolution}

We can use DFDs and chronological reference tables to record changes in sociotechnical systems and their VASs.
Table~\ref{table:changes}, for example, documents the changes for the school choice system in Boston. The matching mechanism (process 2) was the first process to automate, and then update its algorithm on three separate occasions~\cite{abdulkadirouglu2005boston, abdulkadiroglu2006changing, mcdermott2018quality}. Moreover, \citet{wilson2009smartchoice} suggested adding a decision support system for students based on predictions of their development in different schools (process 1) as 
an alternative to other offline processes (e.g., advice from counselors and anecdotic examples based on other students' experiences). Based on Table~\ref{table:changes}, one can easily reconstruct the data-flow diagram in Figure~\ref{fig:rs+mm} depicting the system at different times.

Accompanying DFDs with chronological reference tables documenting VAS changes 
is important for several reasons. First, the literature introducing changes also motivates them. Thus, this literature 
provides important learnings that could be transferred to new contexts (e.g., countries) implementing similar VASs. For example, Boston Public Schools (BPS) changed the mechanism from Boston to Deferred Acceptance (DA) because the first was not strategy-proof, i.e., students could be strategic and misreport their preferences. Policymakers argued that non strategy-proof mechanisms are unfair as they ``provide an advantage to families who have the time, resources and knowledge to conduct the necessary research"~\cite{abdulkadiroglu2006changing}. This motivated the later proposal for using DA in the Chile school choice setting~\cite{correa2019school}.
Second, such a concise and transparent documentation of system design helps understand on which version of the systems prior audits, data analysis, or surveys were conducted. This is key for reasoning on whether results still apply or a new analysis is needed. 
Third, Table~\ref{table:changes} shows the algorithms currently in use, making them easy to incorporate into more realistic agent-based models.

\begin{table}[t]
\centering
\small
\begin{tabular}{p{.75cm}|p{6.75cm}}
    Year & Change \\
    \hline
    \hline
    1987 & Decision of BPS to adopt a choice-based assignment plan (process 2). 
    \\ \hline
    1999 & BPS adopt the Boston Mechanism (for process 2). \cite{abdulkadirouglu2005boston}\\ \hline
    2001 & The No Child Left Behind Act of 2001 allows students at low-performing schools (failing assessment goals in two consecutive years) to change institutions.\\ \hline
    2005 & Boston's School Committee changed the matching mechanism in use (algorithm for process 2, from Boston to Deferred Acceptance) to remove the need for strategic reporting by parents and improve fairness. \cite{abdulkadiroglu2006changing}\\ \hline
    2008 & Introduces process 1 (prediction alg.): SmartChoice is developed and deployed for a focus group in Charlotte-Mecklenburg \cite{wilson2009smartchoice}\\ \hline
    2013 & Boston Public Schools changes process 2 to the Home-Based Student Assignment Policy. \cite{mcdermott2018quality}\\ \hline
\end{tabular}
\caption{Timeline documenting changes in school choice; for simplicity, it focuses on Boston Public Schools (BPS).}
\label{table:changes}
\end{table}

\subsection{Transfer Learnings to Other Application Domains}
Outcome-driven allocations are also used in other matching markets such as labour market~\cite{paparrizos2011machine}, refugee assignment~\cite{bansak2018improving}, and course allocation~\cite{kurniadi2019proposed}.
There are, however, differences. For example, while the school choice VAS presents recommendations to students, in refugee assignment, it solves a constrained optimization problem before suggesting an allocation to placement officers~\cite{bansak2018improving}. 
Despite this difference, the two settings are both susceptible to adversarial interaction attacks~\cite{ionescu2023strategic}, demonstrating that learnings about one VAS could transfer between domains. We, thus, believe harmonized representations and documentations of VAS designs across applications can facilitate fast transfers of learnings. 

\section{Conclusion}


Our work introduces a formal framework for visibility allocation systems (VASs), by defining both VASs and their principal comprising tools. It also discusses how to use data flow diagrams (DFDs) to represent sociotechnical systems relying on VASs and support the evaluation of the long-term effects of these VASs. Using school choice as an example, we demonstrate how our framework can unmask hidden interactions, support documenting findings, and system design changes. Ultimately, the framework helps connect prior results from different research fields under one model and reason about the societal impact of design choices. 

Altogether, we believe our work can support contemporary research and legislative efforts for increasing transparency, promoting understanding for market operators, and developing a harmonized standard for assessing and managing risks across stakeholders~\cite{EU_AI_Act_2024}. We also envision it as a first step toward building an integrated simulation environment for VAS-based sociotechnical systems, where researchers and developers in different fields can independently contribute with their own implementation (e.g., for recommender systems~\cite{graham2019microsoft} or human behavior in response to VAS-outputs~\cite{schweitzer2010agent, muric2022large}) toward analysing the risks of VASs and their alignment with human values. As proved by other domains~\cite{varga2008overview, hines2022neuron, luke2005mason}, building such an environment could accelerate VAS evaluation, getting us closer to matching their fast-paced development and real-world deployment\footnote{We provide a more detailed discussion within the appendix. To demonstrate the feasibility of using our framework for building a simulation environment, we created a Python interface that enables users to define dataflows, tools, and metrics: https://github.com/deutranium/Visibility-Allocation-System. 
}.


\bibliography{VAS}

\section{Appendix}

This appendix contains additional information on several topics covered within the main text. First, we extend the literature and explore more connections between visibility allocation systems (VASs) and prior work. Second, we provide associated problem formulations for and describe the role of other, more complex tools encompassed within VASs, namely search, recommender systems, moderation, and forecasting. The third section provides details on several different stakeholders usually impacted by VASs; it also explores metrics relevant for each. Finally, we include a discussion section which considers the practical limitations of managing large system documentation, the connection between data-flow diagrams and other system representations, how this work can be a first step towards building an integrated simulation environment, and how it can support ongoing legislation efforts.

\section{Extended Related Work}
Due to the interdisciplinary nature of VASs, our work connects with multiple lines of research. We briefly review the most important connections in the main text. Below, we provide a more extensive literature review on each topic presented there.

\subsection{Other (semi)-automated systems} 
While a few other previously introduced types of systems bear similarities with the notion of \textit{visibility allocation system} (VAS) we introduced here, they are overall distinct. In Computer Science, \textit{information retrieval} (IR) and \textit{information filtering} (IF)  both aim to support users getting the information they need, but in slightly different ways~\cite{belkin1992information}. IR attempts to select relevant information from a generally static database in order to match queries provided by users who are largely unknown to the system; differently, in IF data is typically dynamic and users interact repeatedly with the system~\cite{hanani2001information}. Together, both are part of \textit{information access systems}, which want to present users with items that satisfy their information need~\cite{ekstrand2022fairness}. A different class of systems is \textit{decision support systems}: technical systems helping decision-makers address unstructured problems~\cite{sprague1980framework}. While information access does not include, for instance, machine learning (ML) models trained to make predictions based on offline interaction data, such models are part of decision support. \textit{Information system} is a more general term, yet it is usually restricted to organizational contexts and includes technical aspects such as hardware~\cite{bourgeois2019information}.

Importantly, other processes that decide what many users see every day and ultimately affect society, such as ad auctions~\cite{ali2019discrimination}, are not part of any of these systems. This is why prior work has to recourse to locally defined, ad-hoc terms, such as 'the algorithm'~\cite{grandinetti2023examining}, semi-autonomous algorithmic technologies~\cite{pajkovic2021algorithms}, systems for content processing and propagation~\cite{narayanan2023understanding}. We cover this gap by introducing VAS, a term encompassing all technical systems that, based on large datasets of alternatives, decide what users see.



\subsection{Societal impact of VASs}
    The tools comprising VASs impact society. 
Machine learning models, for example, can present algorithmic biases and reinforce existing societal inequalities \citep{oneil2016}. Algorithms used for recruitment show bias against women \citep{dastin2018amazon}, and women are also shown fewer ads about high-paying jobs \citep{datta2015automatedexperimentsadprivacy}. More generally, underrepresented groups in the training data lead to a systematic discrimination, as the model tends to learn the biased representations \citep{Jiawei2023}. These societal impacts are often amplified in the long-term effects through feedback loops \citep{pagan2023, mansoury2020}. 

Studies on specific platforms show their broader societal impacts. Social media platforms contributed to the spread of misinformation on topics such as COVID-19, politics, health, science, and finance \citep{Vicario2016,Vosoughi2018}.
Social media are also known to create echo chambers where users are mostly exposed to information or users that have similar beliefs and opinions \citep{Cinelli2021, delvicario2016echo}. As a consequence, users are not shown diverse content, ultimately contributing to polarization, where users become more extreme and share the same ideas within one chamber \citep{Garimella_Weber_2017}. 
Prior work also shows how changes in the algorithms that fall under VASs can shape user behavior and social interactions. For instance, the introduction of the feature 'People You May Know' on Facebook led to a drastic increase of triangles and new connections~\cite{Zignani_2014}. On Netflix, \citet{Malik_Pfeffer_2016} found evidence of discontinuity in user behavior (i.e., an abrupt increase in the ratings). More generally, \citet{brinkmann2023machine} explains how systems of machines influence culture evolution as they generate, select, transmit, and alter cultural artifacts.

While none of these studies speak directly of VASs, they all show that isolated tools and specific platform designs have a major impact on societal structures and behaviors. By introducing VASs and a framework for documenting systemic changes and learnings, we aim to support a deeper understanding of these systems to better anticipate, monitor, and potentially mitigate their negative impact. 

\subsection{Relevant research fields and methodologies}
A variety of research fields are relevant to the study of VASs, each contributing unique insights into the design, impact, and regulation of these systems.
Computer science provides the technical foundation through algorithm development and audits, focusing on system architecture and optimization ~\cite{Ricci2011, Metaxa2021}. 
Economics enriches this understanding by modeling incentives, competition, and strategic interactions via game theory, mechanism design, and agent-based models~\cite{Varian2007}. Behavioral economics and psychology complement this by using experimental and survey methods to examine bounded rationality, individual decision-making, and contextual influences on user behavior~\cite{Kahneman2011, Goldstein2008, Arnott2019}, alongside survey and qualitative approaches~\cite{Groves2009, Braun2006}. Philosophy provides analytical frameworks to define and evaluate normative concepts such as fairness, transparency, and accountability, revealing how VASs inherently encode value judgments within their technical specifications and operational parameters~\cite{Binns2018, Floridi2018}.
Sociology contextualizes these systems within broader social structures using ethnography, discourse analysis, and institutional theory~\cite{Fourcade2016, Ziewitz2016}. Finally, legal scholarship addresses crucial questions of regulatory compliance, liability, and accountability frameworks governing these systems~\cite{Citron2008, Wachter2017}. 
Taken together, these disciplinary approaches illuminate different facets of VASs, ranging from algorithmic design and user behavior to legal and normative implications.

Increasingly, scholars across fields argue that addressing the complex design and impact of what we here call VASs requires an integrated, multidisciplinary approach.
Accordingly, \citet{Starke2025} argue that fairness in AI extends far beyond technical concerns, as it is \textit{'deeply connected to societal values and ethical considerations'}; thus it cannot be understood or addressed in isolation from broader legal, social, and ethical contexts.
\citet{Gerdon2022} emphasize that social sciences are essential to understanding and mitigating structural inequalities in algorithmic decision-making. They outline how such inequalities can emerge from biased data, problematic formalizations of fairness, and unequal interactional dynamics. They argue that these challenges cannot be addressed through technical means alone.
\citet{Cheong2024} focus on transparency and accountability in AI systems, reviewing the legal and ethical challenges involved. The author underlines the need for collaborative frameworks that include legal scholars, ethicists, and data scientists. 
These position papers exemplify a growing consensus: the tools and systems comprising VASs shape and are shaped by societal dynamics - designing, implementing, and evaluating them, therefore, demands integrating insights from multiple disciplines.

Despite recent research efforts to bridge the divide between disciplines~\cite{Stray2024, mackenzie2024, preece1994, Lazer2009, Mitchell2021, felt2016}, fragmentation persists. For example, \citet{Langer2021} observe that research on AI explainability remains scattered across disconnected disciplines, complicating efforts to ensure stakeholder-aligned system design. Similarly, \citet{Lenders2024} call for fairness research to be grounded in robust disciplinary perspectives, particularly technical and legal. They caution that although disciplinary depth is important, it can deepen fragmentation if not paired with active efforts to integrate perspectives.
This ongoing fragmentation is not just a matter of isolated research efforts, but reflects deeper structural challenges: disciplines remain detached from one another, operating under different epistemological assumptions and prioritizing distinct methodological approaches. Our framework helps address these challenges of multidisciplinary collaboration by offering shared representations and definitions that facilitate cross-disciplinary understanding. By decomposing VASs into modular components and providing harmonized terminology, and introducing data-flow diagrams and system-level metrics, we create common ground for dialogue and evaluation. This supports more coherent collaboration across technical, legal, and ethical domains.

\subsection{Legislation on AI Systems}

Recent regulatory efforts converge on a common premise: AI systems must be governed according to the visibility they grant, the risks they pose, and the social contexts they shape. VAS underpins key impactful technologies that are subject to regulation. Key pieces of legislation include EU's Artificial Intelligence Act \cite{EU_AI_Act_2024}, Digital Services Act \cite{EU_Digital_Services_Act_2022} and the General Data Protection Regulation \cite{GDPR}. Additionally, UNESCO's Recommendation on the Ethics of AI \cite{UNESCO_AI_Ethics_2021} as well as US Executive Order 14110 \cite{US_Executive_Order_14110_2023} (revoked in January 2025) have served as influential normative reference points.

Adopted by the European Parliament in March 2024, the AI Act is the world’s first binding piece of AI legislation. It aims at ensuring proper functioning of the EU internal market by placing harmonised rules regarding the development and deployment of AI systems \cite{EU_AI_Act_2024}. Using a risk-based approach, it prohibits eight “unacceptable-risk” practices, including social scoring, biometric categorisation based on sensitive traits, untargeted facial-image scraping and emotion recognition in workplaces or schools.

All high-risk systems—among them VAS used in critical sectors such as education, employment, credit, migration control and democratic processes—must implement a risk-management cycle, high-quality data governance, technical documentation, logging, human oversight and public transparency. For powerful general-purpose AI (GPAI) models, additional duties include systemic-risk evaluation, incident reporting and publication of training-data summaries.

Across jurisdictions, the regulatory trajectory is clear: visibility must be allocatable not only to end-users (through explanations and controls) but also to regulators and affected stakeholders (through audit-ready logs, risk registers and change-management records).

\section{Additional tools}

Search, recommendation systems, moderation and forecasting are four other tools that a VAS can use. We present each, in turn, below. While we consider the tools presented below and in the main text to be the primarily used ones, the list is far from exhaustive, and could be extended to include tools such as ad auctions~\cite{Petropoulos_2022} and query suggestions~\cite{Noble2018}.

\textbf{Search} tools help users find information by returning content in response to a query~\cite{knuth1998art}. Input includes a user-specified search criterion, usually a query, and a set of alternatives to search through. Design choices determine how the tool compares the query to each alternative in the dataset. The output is a subset of alternatives that best match the search criterion, typically presented in order of estimated relevance. Search tools address the following problem:
\begin{quote}
\emph{Given a dataset of alternatives, a user query, return a subset of these alternatives scored by their fitness with the query.}  
\end{quote}
Search tools play a central role in VASs by helping users find information. They are integrated into on-site site search bars, academic databases, and most prominently, web search engines. Among these, Google Search has become the dominant gateway to online information - influencing how and what news, knowledge, products and services billions of users access, thereby shaping public understanding, attention, and even political discourse~\cite{Epstein2015, Bink2022, Pan2007, Lorenz2022}. These tools increasingly satisfy information needs directly within the search engine results page, eliminating the need to navigate to external websites and underscoring their central role in shaping user access to information \cite{Gleason2023}.

\textbf{Recommender Systems} (RSs)
are personalized information filtering systems: they collect information on the preferences of users to improve the user’s experience by presenting content based on their individual tastes \citep{rs_1}.
The input consists of a set of users with their associated attributes (e.g., past interactions, demographics, or behavioral patterns), user needs (either inferred or explicitly expressed), and a set of alternatives (usually referred to as items). Designers choose a recommendation algorithm (e.g., collaborative filtering), thus effectively weighting the criteria used for recommendations. The output is a score for the fitness of each item for a given user. Based on these scores, the system produces a personalized ranked list of recommendations. Formally, RSs address the following problem: 
\begin{quote}
    \emph{Given datasets of alternatives, user information, and an (implicit or explicit) user need, return a subset of alternatives predicted to best satisfy the user need. }
\end{quote}
RSs are used to reduce users' effort and time searching for relevant content in various domains like tourism, e-learning, e-commerce, movies, and news~\cite{rs_2}. They personalize and control which items are visible, making them critical for VASs

\paragraph{Moderation} tools are typically used on online platforms to remove content or exclude users, and can be manual, similar to filtering, or automated \citep{grimmelmann2015virtues}. Automated content moderation aims to identify and regulate content that might be inappropriate or doesn't align with the platform's guidelines \citep{gorwa2020algorithmic}. By applying algorithms to detect hate speech, misinformation, harmful or hateful content, these systems can adjust the visibility of such content and thus modify or suppress it. 
The input consists of a set of alternatives, generated by contributors of the platforms. The design is centered around a moderation function, often based on a policy that defines acceptable content. The moderation function assesses whether an alternative aligns with a given set of content policies. Depending on the outcome of this function, the tool may take actions such as removing the content, or flagging it as sensible.  
\begin{quote}
    \emph{Given a dataset of alternatives, a set of policies, a set of possible moderation actions, determine whether each alternative aligns with the policies and choose the appropriate action. }
\end{quote}
Online platforms and websites host an increasing volume of user-generated content, which can lead to an increasing volume of harmful, false and dangerous such content \citep{NEURIPS2020_1b84c4ce, BOGOLYUBOVA2018151}. Consistent and transparent moderation that can be enforced by different stakeholders is thus a key component of VASs. 

\paragraph{Forecasting}
tools, widely used for decision-making and planning, support visibility decisions by making predictions about the future based on historical knowledge \citep{DEGOOIJER2006443, Petropoulos_2022}. Their input includes past and present data, and sometimes partial information about the near future (e.g., planned events or expected signals). Design choices vary widely and may include statistical models, machine learning algorithms, or domain-specific rules that account for trends, seasonality, uncertainty, or risk. The output consists of predicted future values, often accompanied by risk assessments, confidence scores or probability intervals, offering guidance on likely developments that may inform visibility decisions downstream.

\begin{quote}
    \emph{Given a dataset of historical observations (i.e., pairings of features to outcomes) and a set of new inputs (i.e., features), return predicted future observations for each instance in this set.
    }
\end{quote}
\citet{Petropoulos_2022} present a wide range of forecasting methods and practical examples of their applications. Forecasting tools have been extensively used for resource planning and allocation in, e.g., school admissions, refugee assignment, or medical residency matching~\cite{wilson2009smartchoice,refugees_bansak, nrmp}. In such domains, decision makers can scrutinize alternatives based on their predicted chances of achieving a positive outcome. This can ultimately have a great impact on the individuals about which decisions are being made (e.g., students, refugees). For example, biased datasets or algorithms can disproportionally affect minorities or disadvantaged groups and reinforce inequalities within society~\cite{Mitchell2021}. Moreover, as we show in the case study of this work, these predictions impact the very population they aim to predict; this leads to a feedback loop known as performativity, where predictions can have a causal influence on the decisions of agents that build the datasets of observations~\cite{perdomo2020performative, hardt2025performative, ionescu2021agent}.

\section{Adapting Data Flow Diagrams for Sociotechnical Systems}
Prior work on DFDs focused on modeling process structures for initial business, rather than societal, process analysis and design~\cite{falkenberg1991understanding}. As such, these diagrams often include many administrative processes that have minimal (if any) societal effects. For example, the first DFD presented by \citet{demarco1979structured} models many administrative processes such as \textit{recording payment}, \textit{producing invoice}, and \textit{producing payment confirmation}. It has, however, far less details on the processes that help customers find the relevant trainings. Similarly, the example DFD presented by \citet{cheema2023natural} includes processes for \textit{Amazon account creation}; yet it presents the \textit{browsing of products} as a general process based on an explicit request from the customer to receive product informaiton. As explained by ~\citet{yourdon1979structured}, the detail and focus of the DFD needs to adapt to the problem and goal of the designer. For studying the connection between design choices of VAS and societal effects, we argue for considering the comprising tools (e.g., search, recommender systems, moderation) as the main processes. 


Another difference when using DFDs for the study of sociotechnical systems is that external entities (stakeholders) are no longer simple sources or sinks which are outside the domain of our study~\cite{demarco1979structured}. They take an active part in creating new data and reshaping the VAS. They also interact with each other. As such, they can become part of processes. For example, when a new YouTube video (item) is created, the content creator (producer) might record the video outside, interview other people, discuss news or events. Thus, the creation of videos becomes a process in which both the producer and society take part. We represent such relationships by dotted arrows.


\section{Stakeholders}
Below we enumerate some of the key stakeholders together with the metrics that are frequently used to measure their entities of interest.

\paragraph{Users}
Users are the primary consumers of VASs. They engage with ranked or filtered content to satisfy information needs, make decisions, or participate in social discourse. Their experience of the system is typically guided by personalized recommendations, search results, or curated feeds. Users may have limited awareness and control over the mechanisms producing what they experience. They provide, however, input (e.g., via clicks, queries, and ratings) which is fed back into the system, ultimately influencing future VAS outputs~\cite{Rieder2018, Eslami2015}.
Users may prioritize metrics related to perceived service quality and cost-effectiveness, such as recommendation relevance and user satisfaction scores.

\paragraph{Producers}
Producers include those who design, implement, and supply content to VASs, such as developers and content creators. While developers focus on optimizing algorithmic components, content producers rely on visibility to reach audiences, drive engagement, and generate revenue. Although producers can shape visibility allocation through strategic behavior, they often remain constrained to platform policies and shifting algorithmic logic~\cite{Gillespie2018, Caplan2020, Bucher2012}.
From a producer perspective, metrics may focus on engagement metrics like clicks, conversion rates, and overall visibility of their content, which directly impact their revenue and reach.

\paragraph{Platforms}
Platforms serve as the infrastructure hosting VASs, connecting producers and users and facilitating their interactions within the system. They maintain the technical infrastructure, design the visibility allocation logic, and often govern behavior through moderation and curation, especially in centralized platforms. Moreover, platforms have an influence on who gets seen and also have a stake in being seen themselves, leading to potential conflict of interests~\cite{Bucher2012, Caplan2020, Plantin2018}.
Platforms emphasize system performance metrics such as accuracy, F1 scores, user satisfaction, usability and profitability, which are crucial for optimizing operational efficiency and market competitiveness.


\paragraph{Additional Stakeholders} 
Additional stakeholders include the general public, regulators, and the natural environment. While not directly involved in interactions with VASs, these actors are affected by their broader societal and ecological impacts. Systemic effects of VASs on additional stakeholder include the polarization of public discourse, unequal access to critical information (e.g., health or political content), and the spread of misinformation. VASs may also reinforce societal biases or contribute to cultural homogenization by privileging dominant perspectives~\cite{Noble2018, Tufekci2015, Pariser2011, Benkler2018}. Moreover, the environmental footprint of operating large-scale algorithmic systems, from energy-intensive data centers to device obsolescence, affects planetary sustainability~\cite{Crawford2021, Markelius2024, Berthelot2024}. These externalities highlight the need for evaluation frameworks that go beyond user- and platform-centric metrics to account for societal risks, environmental costs, and broader patterns of knowledge diffusion and market dynamics.

\section{Discussion}

\paragraph{Managing large system documentation.} As with any modeling technique, there is a natural tradeoff between simplicity and completeness. Out of the box, data-flow diagrams (DFDs) only partially manage this by having different levels of abstraction. For example, \citet{demarco2011structured} notes in a later edition of his original text promoting DFDs that, although project teams still regard them as useful, these DFDs can take up an entire wall of a war room and become hard to maintain. However, recent collaborative tools such as \citet{miro2025, drawio2025, smartdraw_general2025} can support more tractable ways to work on, maintain, and document DFDs. The resulting diagrams could be made interactive (e.g., by clicking on a process to expand it into a lower-level of abstraction DFD).
Moreover, \citet{cheema2023natural} developed an interface using natural-language processing and rule-based algorithms to draw data flow diagrams. All these tools can help maintain an overview and detailed documentation about large sociotechnical systems, thus improving their transparency.

\paragraph{Beyond data-flow diagrams.} While DFDs are one of the cornerstone methodologies in System Analysis and Design, this field now also uses Object-oriented methodologies and the Unified Model Language (UML)~\cite{siau2022information}. The latter includes for example class diagrams, which provide many additional details that can support system implementation into an executable version. Despite the advantages of these more recent approaches, we opted for DFDs here for several reasons. First, focusing on how data is transformed by tools and processes facilitates evaluations. For example, disparate impact is measured directly on data sets~\cite{feldman2015certifying}. Moreover, \citet{alshareef2021transforming} provide an algorithm for transforming regular DFDs into Privacy-Aware DFDs by inserting privacy checks. Second, DFDs require less notation, thus making them more accessible across expertise. This aspect is crucial for facilitating the interdisciplinary work needed for the study of sociotechnical systems~\cite{Gerdon2022}. Third, requiring companies to provide the detailed specifications typical of UML diagrams can be impractical, as it may expose proprietary implementation details and undermine competitive advantage~\cite{Atif2011, Rothenberg2009}. Finally, DFDs can be integrated with other diagram representations to provide additional insight~\cite{kim2000we, aleryani2024analyzing}. This includes the use of abstraction hierarchy which can represent sociotechnical systems at different levels of abstractions and ultimately reveal the paths between VASs and various system objectives~\cite{joseph2022qualitative}.

\paragraph{Towards an integrated simulation environment.} We also envision this work as a first step toward building an integrated simulation environment for VAS-based sociotechnical systems. Many research fields extensively rely on such environments to analyze, test, and compare alternative settings in a fast, risk-free, and reproducible manner. Examples include OMNeT++, used for communication networks, multiprocessors, and other distributed or parallel systems~\cite{varga2008overview}, NEURON - for biological models of neuronal networks~\cite{hines2022neuron}, and MASON - a toolkit for general multi-agent systems~\cite{luke2005mason}. 

By defining the visibility allocation problem we also specified the scope of such a simulation environment. Through decomposing this problem into tool-specific sub-problems we provide a unifying standard for the inputs and outputs of the tools; as a result, we can chain different implementations of these tools to test alternative VAS designs. These specifications can also serve as guidelines for developing new or adapting existing implementations of processes. As a result, researchers and developers in different fields can independently contribute with their own implementation (e.g., for recommender systems~\cite{graham2019microsoft} or human behavior in response to VAS-outputs~\cite{schweitzer2010agent, muric2022large}) toward the final analysis. Data-flow diagrams comprehensively describe the alternative designs being evaluated. Altogether, we believe that by continuing this line of work with building such an integrated simulation environment, we can accelerate VAS evaluation, getting closer to matching its fast-paced development and deployment of new tools\footnote{To demonstrate the feasibility of doing so, we have developed a Python interface that enables users to define dataflow, tools, and metrics: https://github.com/deutranium/Visibility-Allocation-System. }.  

\paragraph{In service of policymakers.}

We believe our work may help public servants such as regulators and national market authorities to decompose sources of impact of data-dependent algorithmic systems. For instance, our work helps organizations \textit{locate their obligations} by decomposing a system into data, model, policy, and interface layers; VASs pinpoint where AI-critical artifacts (e.g., risk files) belong which are trackable by using various tools such as diagrams. Additionally, this helps decision makers to \textit{quantify systemic risk} by connecting tool-based and system-based auditable metrics to societal outcomes, like fairness. \textit{Unifying the language} of key concepts and relationships can help policymakers assess and audit risks of tools related to VASs, which hopefully contributes to developing harmonised standards.
We hope our framework can be in service of regulators and national market authorities in their work by providing a new set of tools. We recommend policymakers to employ our proposed diagrams and documentation in their evaluation of impacts in VAS-related high-risk systems.

\end{document}